# Concept and Construction of Group Signature with self-proof capacity for confirming and denying

CHENG Xiao-gang[1, 2], GUO Ren[3]
1. College of Computer Science and Technology, Huaqiao University, Xiamen China, 361021
2. Xiamen Key Laboratory of Data Security and Blockchain Technology, Xiamen China, 361021
3. College of Business Administration, Huaqiao University, Quanzhou China, 362021

**Abstract:** With privacy-preserving and traceability properties, group signature is a cryptosystem with central role in cryptography. And there are lots of application scenarios. A new extension concept of group signature is presented, namely group signature with self-proof capacity. For a legitimate group signature, the real signer can prove that the signature is indeed signed by him/her. While for the other members of the group, they can prove that the signature is not signed by him/her. The former can be used for claiming money reward from the police, while the latter can be used for proving one's innocent in a criminal investigation.

## 1. Introduction

Group Signature GS (Group Signature) is a central cryptographic system [1] because it has good characteristics of anonymity and traceability; Any member of the group signature can generate a legitimate group signature, but the ordinary verifier can only verify the legitimacy of the group signature during verification, but cannot find out the specific signer, thereby protecting the privacy and anonymity of the signer, and in the event of a dispute, the group administrator GM (Group Manager) holding the group private key can open the signature to find the signer and achieve traceability.

There are many extensions to group signatures, such as revocable group signature[2], i.e., GM can revoke the signing ability of a member; Group signature with hierarchical revocation [3,4], that is, GM can choose to revoke the signing ability of all members of a sub-unit at the time of revocation; The anonymity of ordinary group signatures is fixed, that is, they are anonymous within all members of the group, while in the hierarchical anonymous group signature [5], members can choose the degree of anonymity when generating group signatures; In K+L conditional group signatures[6], traceability is closely related to the number of signatures generated by a member, that is, the identity of the member who misuse group signatures will be exposed, thereby limiting the number of group signatures generated by a member; The revocable group signature scheme without backward correlation [7] means that even if the current key is leaked, it will not affect the security of the previously generated group signature; Quantum computing, especially the Shor algorithm for large integer factorization [8], poses severe challenges to the current commonly used cryptosystems such as RSA and discrete logarithms, so an important direction of group signature research is to resist quantum attacks, such as lattice-based or encoding-based group signature schemes [9,10,11,12,13,14]; On the other hand, based on the assumption of quantum mechanics, an unconditionally secure quantum group signature scheme can be constructed [15,16]. The central idea of blockchain is decentralization, based on which a centralless group signature scheme can be built [17].

Group signatures also have a wide range of applications, such as tracing the identity of close contacts during the pandemic while protecting privacy [18]; Endorsement in blockchain applications [19], as well as in-vehicle ad hoc networks [20], Internet of Things [21], etc.

This paper proposes another group signature extension, that is, a member can prove whether or not a signature is signed by himself. I.e., the real signer can prove that the signature is self-signed (usually in the form of zero-knowledge proof); Non-signers in the group can prove that the signature was not signed by them.

Related to this article is the Deniable Group Signature[13,22], i.e., with the assistance of the group administrator, non-signers can prove that a signature was not signed by themselves; The main differences from this article are: first, the scheme in this article supports both yes and no proof; The second is that if the group signature certificate is completely anonymous, it is inevitable to require the assistance of the group administrator, otherwise it violates the security of anonymity and untraceability, and the certificate scheme constructed in this article does not require the assistance of the group administrator, at the cost of sacrificing part of the anonymity, that is, the signature generated by the same signer is linkable.

The typical application scenario of this scenario is as follows: in order to investigate a crime or hunt down a criminal suspect, the police issue a bounty, and many people may provide information, apparently before the criminal is caught, the information provider wants to remain anonymous, that is, you can sign the message provided to the police with a group signature. After the criminal is caught, the police publish which information is the most valuable, of course, the provider of the non-valuable information remains anonymous, while the provider of the most useful information can prove that it is the clue provided by himself, so as to receive the reward; On the other hand, for false information published on the Internet, non-signatories can prove their innocence during police investigations, that is, prove that they are not the publisher of the news, that is, the relevant signatures are not signed by themselves.

The construction idea of "Yes-proof" is relatively simple, as long as NIZK (Non-Interactive Zero Knowledge) proof is carried out according to the private or random information held by the signer at the time of generating the signature; The idea of negative proof is to generate a legal group signature for a new random message, and the public key used is different from the group signature to be denied, so as to prove that it is not the real signer of the signature (of course, this can be done only if each signer has only one legitimate group signature private key and corresponding public key).

This article is arranged as follows, the definition and preliminary knowledge of our scheme are given in the second section; The third section is the specific construction of our group signature scheme with self-certification function; Its safety is analyzed and proven in Section IV; A summary and further directions of study are presented in section V.

## 2. Preliminaries

**Definition 1.** Group Signature with Self-Proof Capability for Confirming and Denying (GS-SPCCD):

1) **Setup**: GM sets the group signature parameters, generates the group public key PK (for group signature verification) and the group private key SK ($SK_{join}$ is used to generate the member private key, and $SK_{open}$ is used to open the group signature);

2) **Join**: Member $U_i$ apply to join the group, GM verifies his/her identity, and after passing, GM use the group private key $SK_{join}$ to generates the member's private key $S_i$ and sends it to the member $U_i$;

3) **Sign**: Member $U_i$ can generate group signature δ for any message m using the member's private key obtained from the GM, $S_i$;

4) **Verify**: Anyone can use the group public key PK to verify the legitimacy of a message and group signature pair (m, δ);

5) **Open**: For disputed group signatures, GM can open group signatures to find out the real signer;

6) **Confirming Proof**: For a legitimate group signature (m,δ), the genuine signer can prove that the signature was signed by himself;

7) **Denying Proof**: For a legitimate group signature (m,δ), any member of the group can prove that the signature is not signed by itself, except the real signer;

GS-SPCCD security requirements:

First of all, we have the security requirements of ordinary group signatures, that is, anonymity and unforgeability, and then the security requirements of self-certification function:

1) Perjury, that is, the true signatory cannot prove that he is not the signer, and vice versa, the non-signatory cannot prove that he is the signer;

2) The ability to prove and subsequent signatures, that is, after issuing the certificate of yes or no, it should not affect the member's ability to generate signatures normally in the future;

3) Denial and anonymity and traceability: If you directly have the ability to verify or denial, it is obvious that one or several members can conspire to narrow the scope of the group to which the signatory belongs, violating the anonymity of the group; Therefore, the verification is normally with the help of the GM. This article considers the verification without the help of the GM, which of course amounts to a relaxation of anonymity security requirements.

**Signature of Knowledge**: that is, the signature of message m, a non-interactive zero-knowledge proof of having a certain knowledge (such as discrete logarithm, large integer factorization, etc.), usually expressed as:

$$SPK\{x: \ x \ is \ the \ witness \ of \ a \ certain \ hard \ problem\}(m)$$

where m is the signed message, x is the secret knowledge (i.e., the signing private key), and the mathematical puzzle is the public key corresponding to the signature scheme. For example, the famous Schnorr signature scheme based on discrete logarithm and ROM models, the public key is {g; h; p}, the private key is x, satisfies the relationship：

$$h = g^x \ mod \ p$$

The signature is：
$$SPK\{x: h = g^x \bmod p\}(m)$$
Which can be constructed as follows：
$$(R; s) = \{g^r \bmod p; \ r + H(m||R||h) * x\}$$
Verification equation：
$$g^s = R * h^{H(m||R||h)}$$
where r is a random number and H is a hash function that acts as a random oracle (ROM).

The zero-knowledge feature is that the signature verifier cannot obtain any information about secret x after obtaining the signature, because he can calculate (R,s) by himself as follows: take random numbers s and c, and calculate：
$$R = g^s/h^c$$
Then let the value of the ROM hash function $H(m||R||h) = c$. The resulting (R,s) obviously satisfies the above verification equation, and the signature is legitimate. That is, the signature does not disclose any information about the private key of the signature.

The completeness lies in the existence of a Knowledge Extractor, which can use the Rewind technique to extract the secret x:
$$s_1 = r + h_1 * x$$
$$s_2 = r + h_2 * x$$
That is, let the signer sign two different messages with the same R to obtain the above two linear equations, in which r, x is unknown, and the others are known, so that x can be found, so as to prove that the signer has the signing private key, that is, unforgeability.

## 3. Our construction of GS-SPCCD
### 1) General construction idea

First of all, the general idea of group signature construction is: GM generates the group public key PK and members join private key $SK_{join}$ and open private key $SK_{open}$ of group signature; When a member $U_i$ apply to join a group, GM uses the $SK_{join}$ to generate signature $(U_i, S_i)$ as a certificate to issue to the member, and the group signature is a zero-knowledge proof that the member has a certificate：
$$NIZK\{(U_i, S_i): Verify\_PK(U_i, S_i) = 1\} \ (m)$$
When opening the signature, the GM uses the $SK_{open}$ to open the commitment of the $U_i$ in NIZK to get the $U_i$ plaintext, so as to find out the real signer.

Now consider how to self-certify, the proof is relatively simple, the real signatory publishes the random number R used in the $U_i$ commitment, so as to prove that the commitment in the above NIZK is self-generated, that is, the signature is made by himself; Note that the ID identity information of the member $U_i$ is published, but not the signature $S_i$, so it does not affect its subsequent signing ability.

Self-proof is obviously an NP problem (because the private key and the random number used in the signature are NP-witness), so in theory, there is a non-interactive zero-knowledge proof NIZK to self-proof, and at the same time will not disclose the private key information, that is, self-certification will not affect the subsequent signature security; Then the corresponding "no-proof" is a CO-NP problem, and there is no general NIZK to prove. In addition, if the non-signatory party can generate no-proof, that is, it proves that a signature is not signed by itself, to a certain extent, it also violates the security feature of the

"anonymity" of group signatures, that is, one or several non-signatories can narrow the scope of the real signatories of a group signature, that is, partially traceable. Therefore, in the construction of the aforementioned "deniable group signature" [13,20], the cooperation of GM is required to realize the negative proof.

In this article, we consider another way of "no-proof" that does not require GM cooperation, so it will be more convenient in practical applications, but the disadvantage is that it sacrifices part of anonymity, that is, the signatures generated by the same member are anonymous but linkable, that is, anyone can identify the group signatures made by the same member. As mentioned earlier, this is a relaxation of security that is inevitable, otherwise it violates the security definition of group signatures.

Our construction works as follows：

**Setup**：GM generates a random RSA modulo n, and a random multivariate polynomial modulo n：

$$a_0 + a_1 x^r + a_2 y^s + a_3 z^t \ldots \ldots = 0 \ mod \ n$$

The group public key is n and this polynomial, and the GM private key is, of course, the factors p and q of n；

**Join**：When a member asks to join the group, the GM first verifies the member province and randomly generates a solution to the above public key polynomial $(x_i, y_i, z_i, \ldots\ldots)$ if it agrees to join Sent to this member as the member's private key;

**Sign**：$U_i$ first generate and publish $(X_i, Y_i, Z_i, \ldots \ldots)$：

$$X_i = x_i^r, Y_i = y_i^s, Z_i = z_i^t \ldots \ldots$$

Obviously：

$$a_0 + a_1 X_i + a_2 Y_i + a_3 Z_i \ldots \ldots = 0 \ mod \ n$$

Then for a message m, generate SPK as follows:

$$SPK\{(x_i, y_i, z_i, \ldots..): X_i = x_i^r, Y_i = y_i^s, Z_i = z_i^t, \ldots \ldots\}(m)$$

The result signature is $(X_i, Y_i, Z_i, \ldots \ldots)$ and the above SPK。

**Confirming Proof**：Real signer $U_i$ can generate a valid signature for a new random message $m'$：

$$SPK\{(x_i, y_i, z_i, \ldots..): X_i = x_i^r, Y_i = y_i^s, Z_i = z_i^t, \ldots \ldots\}(m')$$

That is, zero-knowledge proves that they have the corresponding private key of the signature, thereby proving that they are the real signatory.

**Denying Proof**：Non-signers can publish their signature public key $(X_j, Y_j, Z_j, \ldots \ldots)(j \neq i)$，and generate a knowledge signature for a random message $m''$：

$$SPK\{(x_j, y_j, z_j, \ldots..): X_j = x_j^r, Y_j = y_j^s, Z_j = z_j^t, \ldots \ldots\}(m''), j \neq i$$

That is, prove that the public private key of your signature is different from the signing public and private key used in the signature, thereby proving that he is not the real signer.

## 2）Concrete construction example

GM choose RSA modulus $n = 11 * 17 = 187$，Generate a random public key polynomial (note that the exponential parts of the polynomial $r, s, t$ should be coprime with $\varphi(n)$)：

$$3x^3 + 7y^{13} + 11z^7 + 19 = 0 \ mod \ 187$$

$\varphi(n) = 10 * 16 = 160$，Therefore, the selected exponential part must be coprime with it, such as 3, 13, 7 in the above equation, and the inverse of its modulo $\varphi(n)$ are 107, 37, 23, respectively.

The equation after linearization is:
$$3X + 7Y + 12Z + 19 = 0 \mod 187$$

**Join**: The member $U_i$ apply to the GM to join the group, the GM verifies the identity of the member, and if agrees, generates random $X_i, Y_i, Z_i$ that satisfies the above linear equation, and then uses the factorization p, q of RSA modulo n to calculate $X_i^{1/3}, Y_i^{1/13}, Z_i^{1/7}$, that is, the private key $x_i, y_i, z_i$, and then send the member's public and private keys to the member $U_i$. For example:
$$X_i = 112, Y_i = 87, Z_i = 169$$

By the factors of n, the private keys are:
$$X_i^{\frac{1}{3}} = X_i^{107} = x_i = 139$$

$$Y_i^{\frac{1}{13}} = Y_i^{37} = y_i = 32$$

$$Z_i^{\frac{1}{7}} = Z_i^{23} = z_i = 152$$

Which satisfy the public key polynomial:
$$3 * 139^3 + 7 * 32^{13} + 11 * 152^7 + 19 = 0 \mod 187$$

**Sign**: First publish $(X_i, Y_i, Z_i) = (112, 87, 169)$, A knowledge signature is then generated for the message m to be signed:
$$SPK\{(x_i, y_i, z_i): X_i = x_i^3; Y_i = y_i^{13}; Z_i = z_i^7\}(m)$$

For $SPK\{x: X = x^i \mod N\}(m)$, It can be efficiently constructed based on the ROM model as follows, select the random number $r$, let $T = r^i \mod N$, $c = Hash(m)$, where the hash function Hash is regarded as a random oracle, and then let $t = x^c r \mod N$, then the knowledge signature is $(t, T)$, and the verification equation is:
$$t^i = X^c T \mod N$$

For $SPK\{139: 112 = 139^3 \mod 187\}(m)$, we can choose $r = 25$, then $T = 25^3 \mod 187 = 104$, suppose the output the random oracle is $c = Hash(m) = 31$, then $t = 139^{31} * 25 = 40 * 25 = 65 \mod 187$, so the signature for message $m$ is $(t, T) = (65, 104)$:
$$t^i = 65^3 = 109 \mod 187$$
$$X^C T = 112^{31} * 104 = 109 \mod 187$$

**Verify**: First verify that $(X_i, Y_i, Z_i) = (112, 87, 169)$ in the signature satisfies the linearized public key equation, i.e.:
$$3X_i + 7Y_i + 11Z_i + 19 = 0 \mod 187$$

It can be verified that:
$$3 * 112 + 7 * 87 + 11 * 169 + 19 = 0 \mod 187$$

Then verify whether the SPK is legitimate, that is, first calculate the hash value $c = Hash(m) = 31$, and then verify whether the $(t, T)$ in the signature satisfy: $t^i = X^c T \mod N$, which of course is true, because the above calculation $t^i = X^c T \mod N = 109 \mod 187$; similar verify $SPK\{y: Y = y^s \mod N\}(m)$ and $SPK\{z: Z = z^t \mod N\}(m)$.

**Confirming Proof**: Generate a new signature with the same $(X_i, Y_i, Z_i) = (112, 87, 169)$, because this means that the member has the corresponding private key $(x_i, y_i, z_i) =$

(139,32,152), thus indicating that the signature to be proved is indeed signed by the member;

**Denying Proof** : Generate a legitimate group signature for the message in the above signature, but with a different member public key $(X_j, Y_j, Z_j)$, i.e. :
$$X_j \neq X_i$$
$$Y_j \neq Y_i$$
$$Z_j \neq Z_i$$

This shows that their group membership certificate and corresponding private key are different from the actual signer who is trying to prove the signature. Of course, the premise is that each group member can only apply for one member certificate, that is, GM is issuing the member certificate and private key to do a good job of identity verification, and limit a member to apply for only one member certificate and corresponding private key. And for any two members $U_i$ and $U_j$:
$$X_i \neq X_j, Y_i \neq Y_j, Z_i \neq Z_j$$

This scheme is relatively simple to prove yes and no, but an obvious disadvantage is that the anonymity is not strong, because obviously all group signatures made by the same member are linkable, because the same member's public key $(X_i, Y_i, Z_i)$ is used for every signature he made.

### 3) Anonymity enhancement

If you want to enhance anonymity, you can encrypt the member's public key each time you generate a group signature :
$$CX_i = X_i h^{r_x}, CG_x = g^{r_x}$$
$$CY_i = Y_i h^{r_y}, CG_y = g^{r_y}$$
$$CZ_i = Z_i h^{r_z}, CG_z = g^{r_z}$$

And GM has the corresponding private key $x$: $h = g^x$, by which he can decrypt these to get $(X_i, Y_i, Z_i)$, for tracing. But now the knowledge signature SPK is more complicated:
$$SPK\{(x_i, y_i, z_i, r_x, r_y, r_z): CX_i = X_i h^{r_x}, CG_x = g^{r_x}; CY_i = Y_i h^{r_y}, CG_y = g^{r_y}; CZ_i = Z_i h^{r_z}, CG_z = g^{r_z}; and\ x_i, y_i, z_i\ satify\ the\ public\ key\ polynomial\}(m)$$

For confirming proof, the singer can just show the random numbers $r_x, r_y, r_z$, then anyone can verify whether the following are valid or not :
$$CG_x = g^{r_x}, CG_y = g^{r_y}, CG_z = g^{r_z}$$

For deny proof, help from the GM is needed. GM can presents $h^{r_x}$, and the non-signer can generate a new signature with the same random number $(g^{r_x}, h^{r_x})$, but with a different member public key $(X_j, Y_j, Z_j) \neq (X_i, Y_i, Z_i)$ , as described above, thus proving that the signature is not signed with its own private key.

### 4. Security analysis
### 1) Anonymity and unforgeability :

Anonymity: The $(X_i, Y_i, Z_i)$ in the group signature is meaningful to the GM (GM can query the real name of the corresponding group member according to its own database), but it is a random number to others and does not expose the signer's province, but obviously that the different group signatures of the one member are linkable, because the same member's public key $(X_i, Y_i, Z_i)$ is used each time.

Unforgeability: First of all, it is relatively simple to construct $(X, Y, Z)$ which satisfy the linearized public key polynomial, for example we can choose random $(X, Y)$, and then solving a linear equation to get Z, but if you want to get $(x, y, z)$ which satisfy:
$$X = x^r, Y = y^s, Z = z^t \ mod \ N$$
Is impossible based on RSA assumption. Of course, an adversary can first select the random number $r, s$, calculate $X = x^r, Y = y^s$, and then bring in the public key polynomial to solve Z, and then try to solve the corresponding z, so as to obtain the member's public and private key; However, note that according to the assumption that RSA is random permutation, the X and Y obtained in this way are random values, so Z is also a random value, and according to the RSA assumption, it is impossible to find $Z^{1/t}$ for a random value Z.

2) Security of confirming and denying proof

1) The security of the confirming proof (that is, the non-signer cannot generate confirming proof): because the proof must be the same as the original signature $(X_i, Y_i, Z_i)$ to generate a new signature, that is, the certifier must know the corresponding private key $(x_i, y_i, z_i)$ to generate the signature, I.e.:
$$X_i^{1/r}, \ Y_i^{1/s}, \ Z_i^{1/t}$$

This is impossible based on RSA assumption.

2) Security of the denying proof: Because when generating a denying proof, the certifier must use a new signature with a new member's public key $(X_j, Y_j, Z_j)$ different from the original signature, that is, the private key he has is different from the private key used in the original signature; Similar to the above proof, it is relatively simple to generate a random $(X_j, Y_j, Z_j)$ public key polynomial satisfies linearization, but it is impossible to obtain the corresponding signed private key $(x_j, y_j, z_j)$ according to the RSA assumption; Of course, the premise here is that each member can only apply for one member key, that is, GM should strengthen the review process of joining, limiting each member to join just once and obtain only one public-private key pair.

5. Conclusion

This paper proposes a new concept of group signature extension, GS-SPCCD with self-certification function, that is, the real signer of the group signature can prove that the signature of a message is generated by himself, while the non-signer can prove that the signature of a message is not self-signed, and does not require the assistance of GM when generating such a denying proof.

The former can be used to receive the reward of the police for providing criminal clues, while the latter can be used to prove his innocence in the police investigation. A specific and efficient scheme is constructed based on RSA hypothesis, polynomials, etc., but its disadvantage is that the group signature made by the same member is linkable, that is, the anonymity is not strong enough; However, this relaxation of security is also inevitable, otherwise when it is rejected, according to the anonymous security of the group signature, the assistance of GM is inevitably required, which will obviously bring a greater burden to GM in practical applications; In the future, we will continue to study how to construct more efficient GS-SPCCD schemes and how to achieve a better balance between the enhancement of anonymity and the burden of GMs.